\documentclass[prl,aps,amssymb,showpacs,twocolumn]{revtex4}
\usepackage{amsmath}
\usepackage{amssymb}
\usepackage{amsthm}
\usepackage{amsfonts}
\usepackage{listings}
\usepackage{enumerate}
\usepackage{latexsym}
\usepackage[dvips]{graphicx}
\usepackage{psfrag}
\usepackage{bm}
\usepackage{graphicx}
\usepackage{subfigure}

\newcommand{\beq}{\begin{equation}}
\newcommand{\eneq}{\end{equation}}
               
\input{epsf}

\begin{document}

\tolerance 10000

\newcommand{\vk}{{\bf k}}
\def\ns{^{\vphantom{*}}}
\def\ket#1{{|  #1 \rangle}}
\def\bra#1{{\langle #1|}}
\def\braket#1#2{{\langle #1  |   #2 \rangle}}
\def\expect#1#2#3{{\langle #1 |   #2  |  #3 \rangle}}
\def\cH{{\cal H}}
\def\half{\frac{1}{2}}
\def\sut{\textsf{SU}(2)}
\def\suto{\textsf{SU}(2)\ns_1}
\def\kF{\ket{\,{\rm F}\,}}


\title{Non-local order in gapless systems: Entanglement Spectrum in
  Spin Chains}

\author{Ronny Thomale$^{1}$, D.~P. Arovas$^{2}$, and B.~Andrei Bernevig$^{1}$}

\affiliation{$^1$ Department of Physics, Princeton University,
  Princeton, NJ 08544, USA}
\affiliation{$^2$ Department of Physics,
  University of California at San Diego, La Jolla, California 92093,
  USA}

\begin{abstract}
  We show that the entanglement spectrum can be used to define
  non-local order in \emph{gapless} spin systems. We find a gap that
  \emph{fully} separates a series of generic, high `entanglement
  energy' levels, from a \emph{flat} band of levels with specific
  multiplicities that uniquely define the ground-state, and remains
  finite in the thermodynamic limit. We pick the appropriate set of
  quantum numbers, and then partition the system in this space. This
  partition corresponds to a very non-local real-space cut.  Despite
  the fact that the Laughlin state is bulk gapped while the
  antiferromagnetic spin chain state is bulk gapless, we show that the
  $S=\half$ Heisenberg antiferromagnet in one dimension has an
  entanglement spectrum almost identical to that of the Laughlin
  Fractional Quantum Hall state in two dimensions, revealing the
  similar field theory of their low-energy edge and bulk excitations
  respectively. In addition, we investigate the dimerization
  transition from the perspective of entanglement gap scaling.
\end{abstract}

\date{\today}

\pacs{73.43.Â\u2013f, 11.25.Hf}

\maketitle


Gapped topological phases of matter usually lack local order
parameters that can distinguish them from trivial ones.  In the
presence of a gap in the excitation spectrum, several non-local
indicators of the topological nature of the topological phase, such as
ground-state degeneracy on compact high-genus
manifolds~\cite{haldane86prb3844} as well as the structure of edge
modes and their scaling exponents~\cite{wen90prb12844} exist, but do
not fully describe the topological phase. In the absence of an
excitation gap, the nature of quantum order is much less understood,
especially in systems lacking a local order parameter. For example,
torus degeneracy cannot distinguish gapless phases, since the mixing
of infinitesimally low-energy excitations can lead to many new states
quasi-degenerate with the true ground-state.  There are several
examples of gapless phases of matter, most notably in one-dimension
(but also in higher dimensions~\cite{hermele-05prb104404}), which are
known to display highly non-trivial properties.  The parade example is
the spin-$\half$ Heisenberg chain, a spin liquid which exhibits new
elementary excitations, the
spinons~\cite{faddeev-84jsm241,andrei-79prl1698}, which fractionalize
out of the usual antiferromagnetic spin wave.  While for gapped
systems such as the AKLT spin chain, non-local order parameters have
been found, no similar quantity exists for gapless spin-disordered
chains.


The advent of quantum information theory has led to new concepts, such
as quantum entanglement, which have proved useful in further
characterizing quantum order.  The majority of research to date has
concentrated on entanglement entropy of a state $\ket{\psi\ns_0}$.
Specifically, let the Hilbert space for an $N$-body system be written
as a direct product $\cH=\cH\ns_A\otimes\cH\ns_B$.  It is easiest to
imagine real space partitions, where the division corresponds to some
prescribed real space `cut'.  Then $\rho\ns_A\equiv
\textsf{Tr}\ns_B(\rho)$, where $\rho=\ket{\psi\ns_0}\bra{\psi\ns_0}$,
is the reduced density matrix for the $A$ component.  The quantity
$S\ns_A=-\textsf{Tr}\ns_A\,(\rho\ns_A\ln\rho\ns_A)$ is the
entanglement entropy, which provides a measure of the quantum
entanglement of $\ket{\psi\ns_0}$ with respect to the partitioning
$(A,B)$.

Different scalings of this quantity with the size of $A$ for gapped
and gapless systems have been proposed and
proved~\cite{calabrese-04jsmp06002}.  Still, the entanglement entropy
remains a single number, and can only provide a limited
characterization of topological order.  To this end, Li and Haldane
(LH) in a recent insightful paper~\cite{li-08prl010504} proposed and
numerically substantiated that the \emph{entanglement spectrum}, {\it
  i.e.\/} the full set of eigenvalues of $\rho\ns_A$, partitioned with
respect to the equator of the quantum Hall
sphere~\cite{haldane83prl605}, provides a near complete picture of the
topological order in $\nu=\frac{5}{2}$ fractional quantum Hall effect
(FQHE) states.  Writing the eigenvalues of $\rho\ns_A$ as
$e^{-\xi\ns_i}$, where $\xi\ns_i$ is an entanglement level, LH showed
that the low level spectrum for generic gapped $\nu=\frac{5}{2}$
states exhibited a universal structure, related to conformal field
theory, and separated from a non-universal high energy spectrum by an
{\it entanglement gap\/} which was finite in the thermodynamic limit.
This gap itself was proposed as a `fingerprint' of the topological
order present.  It was subsequently shown that the entanglement
spectrum can meaningfully distinguish among states which have similar
finite size overlap with $\ket{\psi\ns_0}$ but different edge
structures~\cite{regnault-09prl016801}.  Other bulk-gapped phases such
as topological insulators have also been shown to reflect their
nontrivial nature in their entanglement
spectra~\cite{fidkowski10prl130502,turner-09cm0909}.

In this Letter, we show how to use the entanglement spectrum to
characterize non-local order in \emph{gapless} systems and phases of
matter, with emphasis on spin-$\half$ spin chains.  We find that
simple real space partitions fail to reveal the nontrivial fundamental
structure in these phases.  Rather, we find the cut must be made in
Fourier (momentum) space. The local cut in momentum space corresponds
to a highly non-local cut in real space.  We only use the
\emph{ground-state} wavefunction of the system and do not assume any
knowledge regarding the gapless excitation spectrum.  If the system is
in the $\suto$ Wess Zumino Witten (WZW) universality class for
spin-$\half$, we show that the low-entanglement level part of the
entanglement spectrum has a special form similar to that of the
Laughlin~\cite{laughlin83prl1395} FQHE state.  The high entanglement
levels are separated by a large entanglement gap from the low
entanglement WZW portion of the spectrum.  This recapitulates the
situation for the ground state of the Coulomb Hamiltonian on the
quantum Hall sphere~\cite{haldane83prl605}, but with two important
differences: (i) we are here dealing with a spin system, and (ii) we
find a {\it complete\/} entanglement gap at all values of the total
momentum, whereas LH found a gap only for a finite range of angular
momentum $L^z$.  We numerically show that the entanglement gap is
finite in the thermodynamic limit for the Heisenberg chain. As one
result of our work, one can classify whether a generic ground-state
spin-$\half$ wavefunction is in the WZW universality class, without
resorting to often unreliable computations of critical exponents.  We also show that the entanglement spectrum is sensitive withe respect to the dimerization transition  - the flow of entanglement levels matches the field theory prediction for the canceling of a marginal operator at exactly the dimerization transition.

\begin{figure}[t]
  \begin{minipage}[l]{0.35\linewidth}
    \psfrag{A}{$\large 1$}
    \psfrag{B}{$\large 2$}
    \psfrag{C}{$\dots$}
    \psfrag{D}{$\large N$}
    \psfrag{F}{$\vert z_i - z_j \vert$}
    \includegraphics[width=\linewidth]{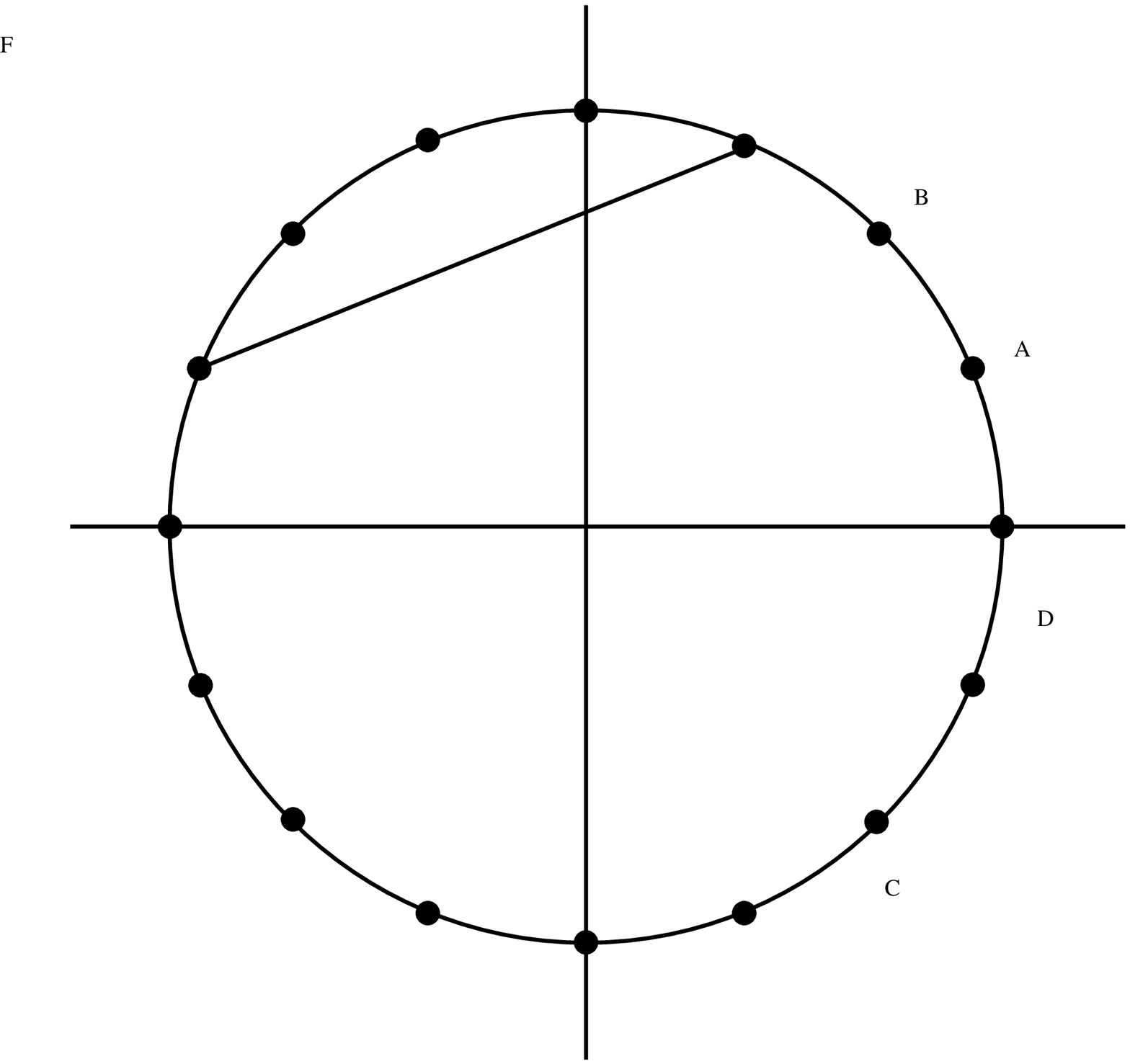}
  \end{minipage}
  \begin{minipage}[l]{0.6\linewidth}
    \includegraphics[width=\linewidth]{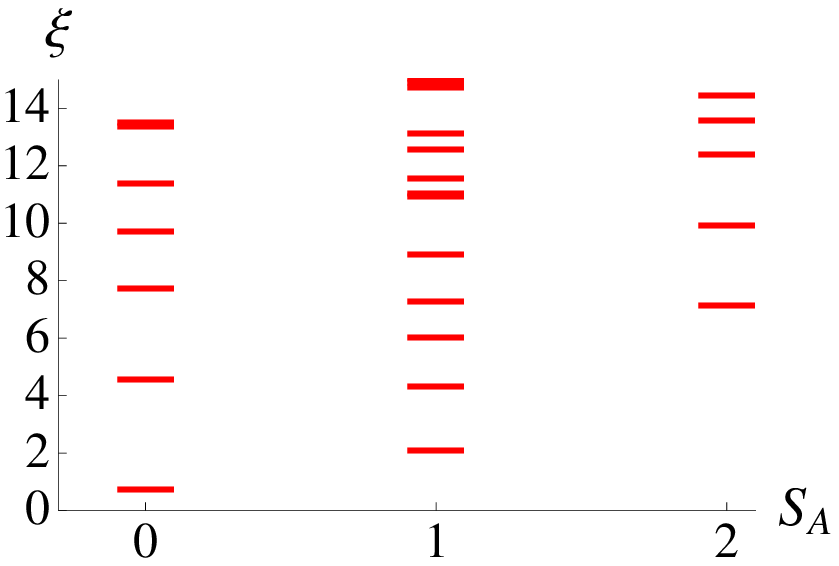}
  \end{minipage}
  \caption{(color online) Left: Unit circle description of a spin chain with PBCs;
    the sites lie at the $N^{\rm th}$ roots of unity.  The chord
    distance specifies the spatial separation of any two sites $z_i$
    and $z_j$.  Right: Typical low-level entanglement spectrum for a
    spatially bisected $N=30$ Heisenberg chain (ground state).
    Entanglement levels along the vertical axis are plotted {\it
      versus\/} total spin $S_A$ quantum number.  Each level has a $2
    S_A+1$ multiplet degeneracy. No immediate structure is visible.}
\label{LatticePlusEnt}
\vspace{-0pt}
\end{figure}

We consider spin-$\half$ spin chains on even-membered rings with
periodic boundary conditions (PBCs).  The $N$ sites are placed on a
circle of radius unity and are hence described by the $N^{\rm th}$
roots of unity: $z\ns_j=e^{2\pi i j/N}\ ;\ j \in \{1,\dots, N\}$ (see
Fig.~\ref{LatticePlusEnt}). Without loss of generality, any ground
state wavefunction of an $\sut$-invariant Hamiltonian can be written
in the form
\begin{equation}
\ket{\Psi\ns_0} = \sum_{j\ns_1,\ldots,j\ns_K} \!\!\!\!\psi(z\ns_{j\ns_1},\ldots,z\ns_{j\ns_K})\,S^-_{j\ns_1}\cdots S^-_{j\ns_K} \, \kF\ ,
\label{groundstate}
\end{equation}
where $\kF=\ket{\!\uparrow\cdots\uparrow\,}$ is the ferromagnetic
state.  The sum extends over all ways to distribute the $K=\half N$
down-spins on the ring, and the weights $\psi(z\ns_{j\ns_1},\ldots,
z\ns_{j\ns_K})$ depend only on the position of the spin $\downarrow$
sites (Holstein-Primakoff representation).  We further presume that
$\ket{\Psi\ns_0}$ is a translationally invariant total spin singlet
(otherwise, in most cases, the state could already be characterized by
some other local order parameter).  In particular, these are also the
conditions for spin liquid states.  Mathematically stated, we require
$\sum_j S^-_j\ket{\Psi\ns_0}=0$ and
\begin{equation}
\psi(z\ns_{j\ns_1+1},\ldots,z\ns_{j\ns_K+1})=e^{iQ}\, \psi(z\ns_{j\ns_1},\ldots,z\ns_{j\ns_K})\ ,
\end{equation}
where $Q=0$ or $\pi$ is the total crystal momentum, depending on
whether $K$ is even or odd, respectively.  Note that the last
condition is a generic property independent of the particular
ground-state wavefunction studied. Spin rotational invariance also
implies that the wavefunction is invariant under interchange of
$\uparrow$ and $\downarrow$ coordinates.

Typically, calculations of entanglement and entanglement entropy
proceed by cutting the system spatially into two disconnected parts
and then computing some combination of the eigenvalues of the reduced
density matrix which decomposes in independent sectors indexed by the
values of the total $S^z$.  If we choose to perform this cut, compute
the eigenvalues of the density matrix, and plot them versus the good
multiplet quantum number $S_A$, one observes an unremarkable structure
such as shown in Fig.~\ref{LatticePlusEnt}. To obtain relevant
information, one must then analyze the distribution of eigenvalues of
the entanglement spectrum~\cite{pollmann-09cm0910} or perform
calculations of the entanglement
entropy~\cite{calabrese-04jsmp06002,fuehringer-08adp922}.  This
suggests that a spatial cut is not the most useful way to understand
the underlying structure of the spin chain, as no differences in the
systems can be suitably resolved~\cite{pollmann-09cm0910}.

\begin{figure*}[t]
  \begin{minipage}[l]{0.90\linewidth}
    \includegraphics[width=\linewidth]{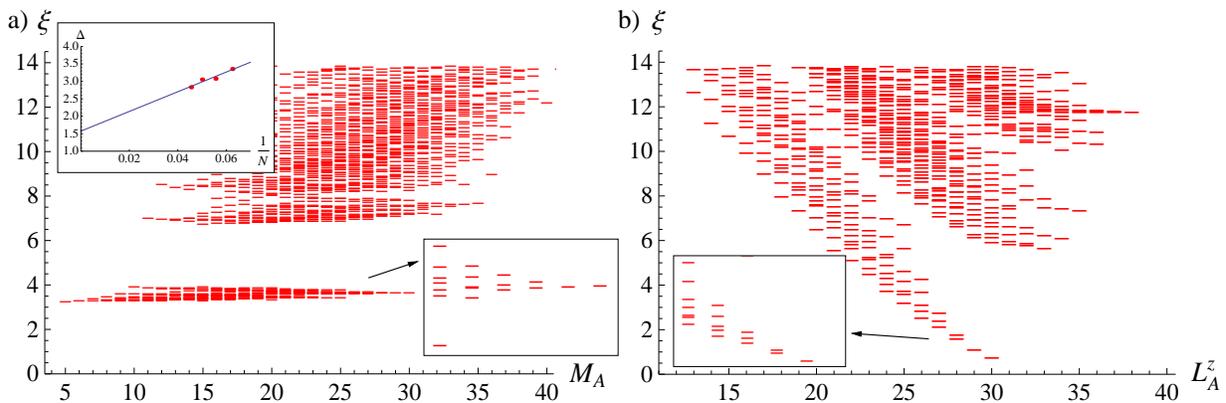}
  \end{minipage}
  \caption{(color online) a) Entanglement spectrum of the Heisenberg model for N=22
    sites, with a cut region containing $5$ magnons. The eigenvalues
    $\xi$ are plotted versus the total momentum subspaces of region
    $A$, $M_{A}$. One observes a low level part well separated from
    higher level contributions. This entanglement gap persists in the
    thermodynamic limit as shown in the upper inset. For the Haldane
    Shastry spin chain, exactly the same low-lying levels are found,
    where the higher levels are shifted to infinity. b) Entanglement
    spectrum of the $K=11$ particle $\nu=1/2$ FQH Coulomb state for a
    half cut on the sphere containing $5$ particles. The counting of
    the low level states is exactly the same. The lower insets in (a)
    and (b) show the analogous level counting $1,1,2,3,5,7$ for the
    two states.}
\label{heisemom}
\vspace{-0pt}
\end{figure*}

However, real space coordinates are only one option of a set of good
quantum numbers in the system.  We can also Fourier transform the spin
operator on each site and go to a magnon-type description. We define
the Fourier transform of the spin flip operators ${\tilde
  S}^-_m=\frac{1}{N} \sum_j z_j^m S_j^-$.  The ground state is then
written as 
\begin{equation}
\ket{\Psi\ns_0}=\sum_{m\ns_1,\ldots,m\ns_K}\!\!\! {\tilde\psi}(m\ns_1,\ldots,m\ns_K)
\,{\tilde S}^-_{m\ns_1}\cdots {\tilde S}^-_{m\ns_K}\,\kF\ ,
\end{equation}
where
\begin{equation}
{\tilde\psi}(m\ns_1,\ldots,m\ns_K)=N^{-K}\!\!\!\!\sum_{j\ns_1,\ldots,j\ns_K} \!\!\!\!
z_{j\ns_1}^{m\ns_1}\cdots z_{j\ns_K}^{m\ns_K}\> \psi(z\ns_{j\ns_1},\ldots,z\ns_{j\ns_K})\ .
\label{fourier}
\end{equation}
We stress that while the momenta $m\ns_i$ also belong to the set
$m\in\{1,\ldots,N\}$, the momentum space spin flips do not behave as
hard core bosons.

The non-orthogonal momentum state basis can thus be represented by
bosonic occupation numbers $n\ns_m$ for crystal momentum $m$.  The
total particle number is $K=\sum_m n\ns_m$, and $e^{iQ}=\prod_m
e^{2\pi i m n\ns_m/N}\equiv e^{2\pi i M/N}$, where~$0 \le n_m \le K\
\forall\ m$.  For generic wavefunctions, there exist multiple total
momentum sectors separated by $2 \pi$, although special spin-chain
wavefunctions can have compact support, {\it i.e.\/} weight in only
one of these sectors.  We normalize the states by the bosonic
normalization of the magnon basis.  Now consider a cut in momentum
along the middle of the momentum orbitals.  Both separate regions $A$
and $B$ can be decomposed with respect to number of particles and
total momentum, which is subject to the constraint $N_A+N_B=K$ and
$M_A+M_B=M$.

Let us first consider the ground state of the $S=\half$ Heisenberg
model, $H=\sum_j \vec{S}_{j} \cdot \vec{S}_{j+1}$, for $N=22$ sites.
For systems of this size, the explicit Fourier transformation
Eq.~\ref{fourier} exceeds numerical limits. Instead, we diagonalize
the spin Hamiltonian directly in momentum space, with matrices
containing up to several billions of scattering elements.  
For the Heisenberg ground state, we find that $98\%$ of the weight of
the state is present in the sector with $M=K^2$. In the thermodynamic
limit, we conjecture that the same low-energy structure of the
entanglement spectrum will be present in all momentum sectors. The
decomposition of the Heisenberg ground state in momentum notation for
a middle cut is presented in Fig.~\ref{heisemom}.  We first observe a
low entanglement level part clearly separated from levels higher up.
For the low-lying states, from right to left, we observe the counting
$1,1,2,3,5,7$ of levels of a $\textsf{U}(1)$ boson mode until we hit a
limit due to finite size of the system. This counting is equivalent to
that for the excitations of the low energy field theory for the
spin-$\half$ Heisenberg chain (the \textsf{SU(2)}$_\textsf{1}$ field
theory is equivalent to that of a free \textsf{U}(1) boson, whose
level counting is equal to the number of partitions of its momentum
above the ground-state).  We then compare to the entanglement spectrum
of the Coulomb FQHE state on the sphere with the same number of
particles.  The $L_A^z$ quantum number on the quantum Hall sphere
corresponds to the total momentum $M_A$ for the cut region,
where the convention is that the momentum counting starts from the
cut, {\it i.e.} for maximum $M_{A}$, the occupied momentum orbitals
have maximal orbital distance to the cut.  This way, equal number of
particles beyond the cut and equal absolute values of $M_{A}$ and
$L_A^z$ specify the same configurations in bosonic notation.  For the
Heisenberg ground state on the chain and the Coulomb state on the
sphere, the entanglement spectra are similar both by level sequence as
well as by state counting, except that in the spin chain case, a
\emph{full} entanglement gap can be defined between all the $\suto$
low entanglement levels and the higher lying ones
(Fig.~\ref{heisemom}).
A finite size scaling analysis of the entanglement gap for the
Heisenberg ground state is shown in the inset of Fig.~\ref{heisemom}a.
Despite the relatively small number of sites that are available to us
on small computers, 
we are confident that the entanglement gap remains
finite in the thermodynamic limit.

\begin{figure}[t]
  \begin{minipage}[l]{0.99\linewidth}
    \includegraphics[width=\linewidth]{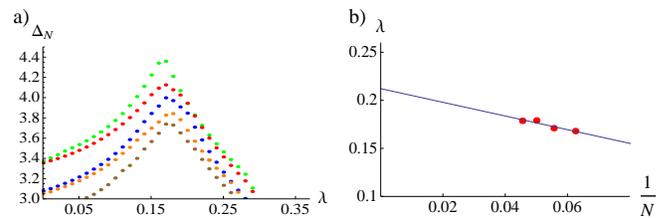}
  \end{minipage}
  \caption{(color online) a) Entanglement gap $\Delta$ versus
    dimerization coupling $\lambda$ for different system sizes. As we
    move away from the Heisenberg point $\lambda=0$, the entanglement
    gap rise, till it finally shrinks again within the dimerized
    phase. b) The dimerization transition is estimated from finite size
    by the $1/N$ scaling of the $\lambda$ with maximum entanglement
    gap for different system sizes. The extrapolation gives
    $\lambda_c^{\infty}\sim 0.215$, i.e. in range of $10\%$ of the
    exact result from CFT.}
\label{dimer}
\vspace{-0pt}
\end{figure}

The entanglement property appears to be intimately linked to the
low-energy properties of the system.  Other regimes of the
spin-$\half$ system possess different signatures of their entanglement
spectrum.  Adding a second-neighbor interaction, $H=\sum_j\vec{S}_j
\cdot \vec{S}_{j+1} + \lambda\,\vec{S}_j \cdot \vec{S}_{j+2}$, we find
that as we shift from the Heisenberg point ($\lambda=0$) to the
Majumdar-Ghosh~\cite{majumdar-69jmp1388} point ($\lambda=\half$),
where the ground state is a dimer crystal, the entanglement spectrum
significantly rearranges, reflecting the failure of the gapless
$\textsf{U}(1)$ boson description of the problem in the dimerized
phase.  From the Sine-Gordon description of the
model~\cite{haldane82prb4925} that we know that the leading
logarithmic operator corrections vanish at the dimerization point.
This corresponds to the vanishing of the Sine-Gordon cosine term at
$\lambda_c$. At this critical point, only subleading operators remain,
and the theory should look more like that of a free-boson than even
the Heisenberg point. We find this is manifested in the entanglement
spectrum. Upon increasing frustration from the Heisenberg point
$\lambda=0$, we find that the lowest (i.e.  leading correction)
high-energy levels first move upwards, which yields an increase of the
entanglement gap $\Delta$ (Fig.~\ref{dimer}a). The dimerization
transition is given by the vanishing of the leading marginal operator
and hence by the maximum entanglement gap. Under this hypothesis, we
obtain a value fo $J_2/J_1=0.215$ in the thermodynamic limit. This is
close to the established $0.24$ value and even more noteworthy as our
largest size system is $22$ sites. As the coupling of the Sine-Gordon cosine term
changes sign at $\lambda_c$, the term is responsible to generate the
dimerization gap.  This is seen in the entanglement spectrum by the
reduction of the entanglement gap, as the dimerization constant
$\lambda$ is tuned across its critical value.
We remark that our numerical effort was heavy: brute-force Fourrier transform of the wavefunction was possible only up to $18$ sites. We then re-formulated our hamiltonian in Holstein-Primakoff basis, linearized, and introduced a term that gives large energy for double boson occupancy on site. We were then able to reach $22$ sites, with non-sparce matrices of   several billion matrix elements.

Beyond the same state counting of the universal low level portions of
their entanglement spectra, there are several other analogies between
the Heisenberg and Laughlin states.  As was first discovered for
quantum Hall trial states~\cite{bernevig-09prl206801}, we likewise
find for the Heisenberg state that, to a large extent, the weights of
the different basis states obey a "product rule" in the sense that the
weights of a Heisenberg state of size $N$ can be composed out of
weight products of sub-parts of the bosonic momentum state
configuration $\{m\ns_1, \ldots m\ns_K\}$, given by Heisenberg states
of accordingly smaller system size.

This strong structural similarity between the Heisenberg state and
Laughlin state is not coincidence. The Heisenberg chain is in the same
universality class as the Haldane-Shastry (HS)
chain~\cite{haldane88prl635,shastry88prl639}, whose ground state
wavefunction is $\Psi(z\ns_1,\dots, z\ns_K)=\prod_{i<j} (z_i-z_j)^2
\prod_i z_i$.  Though of no consequence in real space (where the
entanglement spectra for HS are unremarkable), we found that the
lattice Fourier transformation over all sites (Eq.~\ref{fourier}) of
the HS state yields exactly the weights of the monomials $m_{\{n\ns_K,
  n\ns_{K-1}, \ldots, n\ns_1\}}$ in the $K$-particle Laughlin state.
Hence the entanglement spectrum of the HS ground state, using a {\it
  momentum cut\/}, is {\it identical\/} in its counting to the
entanglement spectrum of the Laughlin state, as shown in
Fig.~\ref{heisemom}~\footnote{The HS state, by virtue of being a
  homogeneous polynomial in the positions of the $\downarrow$ spins,
  always has $100 \%$ weight on the $M=K^2$ sector.  Its component
  configurations in the crystal momentum basis also satisfy a
  `squeezing' property~\cite{bernevig-08prl246802}, just like the
  Laughlin state (this is not obvious: the real-space HS wavefunction
  is written in terms of $K$ coordinates, while the Fourier transform
  sums over all $N=2K$ positions that these coordinates can take on
  the unit circle).  
  Away from the HS model, 
  admixtures of components with $M=K^2\pm j N$ with $j=1,2,\ldots$, as
  well as `un-squeezed' components with $M=K^2$, are generated.  The
  entanglement spectrum is computed taking into account all
  other components but treating the full momentum 
  as a good quantum
  number.}.  We stress that while the entanglement levels will be
different due to the different normalizations for the spin chain and
the QH sphere, the {\it counting\/} is identical.  The HS state is
known to be the conformal gapless phase of all spin models within the
the $\suto$ WZW class, with special quantum group symmetries.  In
terms of the fractionalized spinon excitations, it is the model in
which the spinons are free in the sense that they interact only
through their fractional statistics.  Toward the Heisenberg point, the
spinons get dressed, but retain their fractionality.  The entanglement
spectrum of the HS state consists purely of the universal low spectral
levels of the Heisenberg phase (all the other levels are shifted to
infinity in $\xi$, corresponding to a largely degenerate $O$
eigenvalue level of $\rho_A$). From there, the entering logarithmic
CFT corrections appearing as finite high-energy entanglement levels
can be nicely observed upon interpolating between the ground states of
the HS and the Heisenberg Hamiltonian. This underscores the fact that
the HS and Heisenberg model belong to the same universality class,
with the same relation between the Laughlin state and the Coulomb
state in the FQHE.

%

We have introduced a suitable basis to fruitfully discuss entanglement
spectra of quantum spin chains.  Specifically, we have analyzed the
$S=\half$ Heisenberg ground state wavefunction and observe a universal
set of states separated from higher levels by an entanglement gap
persisting in the thermodynamic limit, bearing strong similarities to
the Laughlin and Coulomb states in the FQHE.  It is hence possible to
determine the non-local structure of a spin system just by examining
the ground state, without any reference to the concrete Hamiltonian,
as we demonstrated for the Heisenberg state in its relation to the
Haldane-Shastry state, as well the dimerization transition from the
Heisenberg to the Majumdar-Ghosh state.  From the present result and
the FQH studies, a unified picture emerges: the low level structure of
the entanglement spectrum reflects the properties of the low-energy
excitations of the system; it is in a one-to-one correspondence with
the elementary excitations of the corresponding field theory.  For FQH
states, which are bulk gapped, an orbital cut as performed
in~\cite{li-08prl010504,regnault-09prl016801} is similar to a spatial
cut due to the localized nature of the Landau orbitals on the sphere.
The entanglement spectrum then reveals the nature of the low-energy
excitations of the gapless edge system.  For Heisenberg spin chains,
the bulk is gapless, but the momentum cut does not longer correspond
to a spatial cut, and the entanglement spectrum reveals the nature of
the bulk gapless excitations. In both cases, these low-energy
excitations are described by $\textsf{U}(1)$ bosonic field theory.

We thank N. Regnault for numerous 
discussions and for allowing us use of his cluster and for numerical help. 
We thank L. Balents for pointing out to us the vanishing leading
logarithmic corrections at $\lambda_c$. We thank F.D.M.  Haldane for numerous
discussions and suggestions. RT was supported by a Feodor Lynen
scholarship of the Humboldt foundation.  DPA is grateful
for the hospitality of the Princeton Center for Theoretical Science.
BAB is supported by an Alfred P. Sloan Fellowship and by Princeton
University start-up funds.



\vfil\eject
\noindent{\bf Supplementary material:}

\begin{figure}[t]
   \begin{minipage}[l]{0.45\linewidth}
     \psfrag{A}{{\small$\xi$}}
     \psfrag{B}{{\small$M^A$}}
     \includegraphics[width=\linewidth]{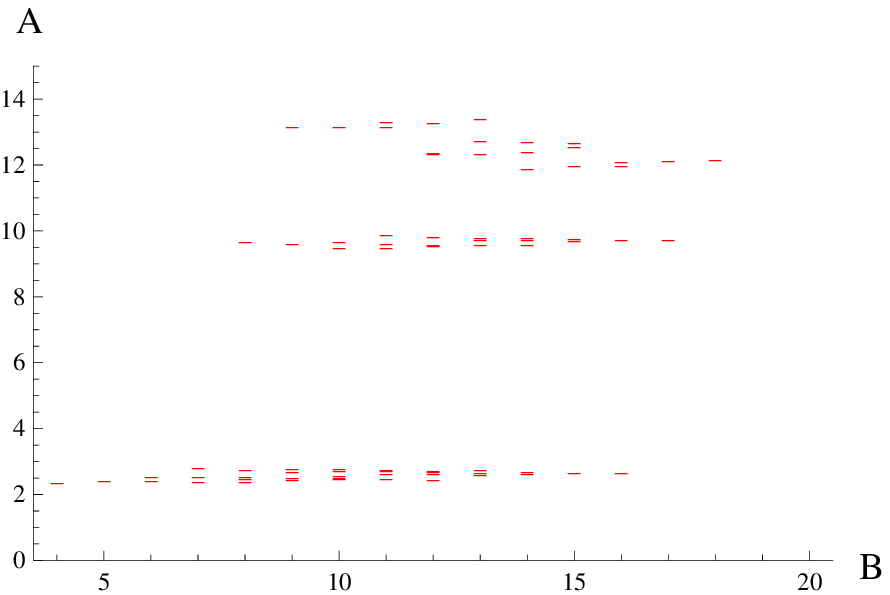}
   \end{minipage}
 \hspace{10pt}
   \begin{minipage}[l]{0.45\linewidth}
     \psfrag{A}{{\small$\xi$}}
     \psfrag{B}{{\small$M^A$}}
     \includegraphics[width=\linewidth]{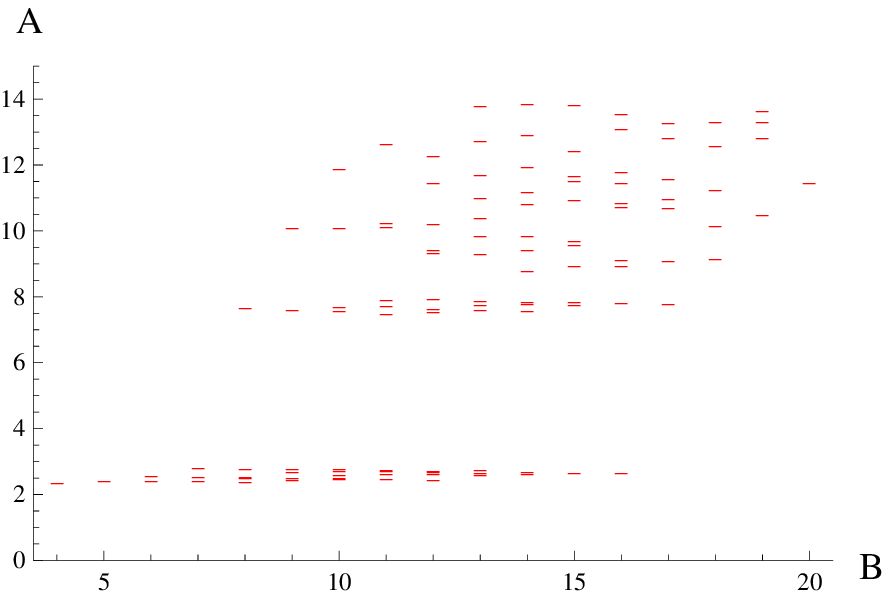}
   \end{minipage}
   \caption{Entanglement gap evolution of the interpolation
     Hamiltonian~\eqref{hint} with $N=16$ sites for the values
     $\gamma=0.75$ and $1.50$.  Note that high entanglement levels
     start to appear immediately away from the HS point, resulting in
     an {\it entanglement gap\/}.  Already at $\gamma=1.5$, the
     spectrum already closely resembles the Heisenberg spectrum.}
 \label{sinhsweep}
 \vspace{-0pt}
 \end{figure}

\emph{(i) Logarithmic corrections from the Haldane-Shastry point.} 
We show how the entanglement spectrum changes from the
 Haldane-Shastry point to the Heisenberg point. This is best
 accomplished by the interpolation Hamiltonian
 \begin{equation}
 H_{\rm int}=J_0 \gamma^2 \sum_{i<j} {\vec{S}_i\cdot \vec{S}_j+\half\over
 2\sinh^{2}(\gamma \vert z_i - z_j \vert /2)}\ , \label{hint}
 \end{equation}
 where the Haldane-Shastry limit is reached for $\gamma \rightarrow
 0$.  From the viewpoint of CFT, as discussed in the manuscript, the
 change in the entanglement spectrum relates to the logarithmic
 corrections entering from the conformally critical HS point.  We find
 that the states enter in a flat band structured fashion corresponding
 to different orders of corrections, as shown in Fig~\ref{sinhsweep}.

\begin{figure}[t]
   \begin{minipage}[l]{0.45\linewidth}
     \psfrag{A}{{\small$\xi$}}
         \psfrag{B}{{\small$L_A^z$}}
     \includegraphics[width=\linewidth]{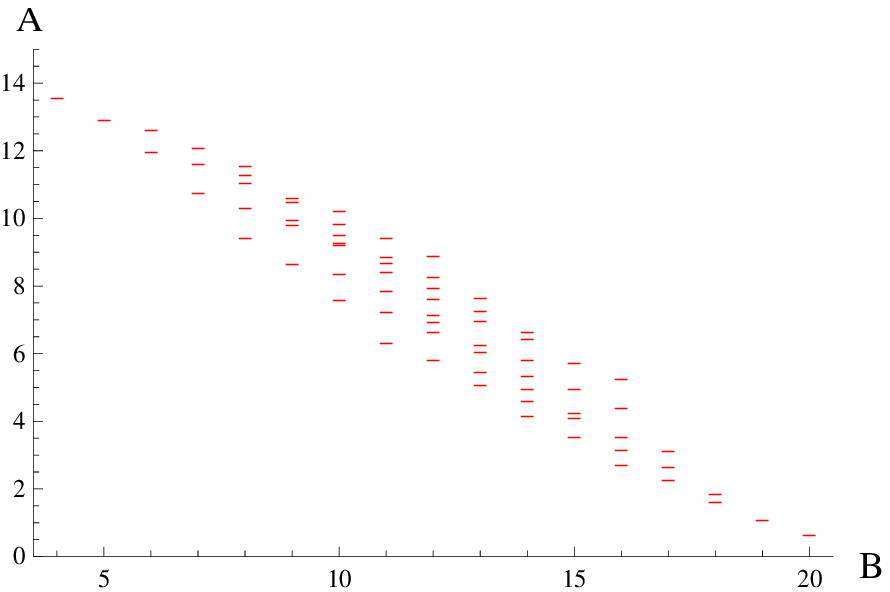}
   \end{minipage}
 \hspace{10pt}
   \begin{minipage}[l]{0.45\linewidth}
     \psfrag{A}{{\small$\small\xi$}}
         \psfrag{B}{{\small $M^A$}}
     \includegraphics[width=\linewidth]{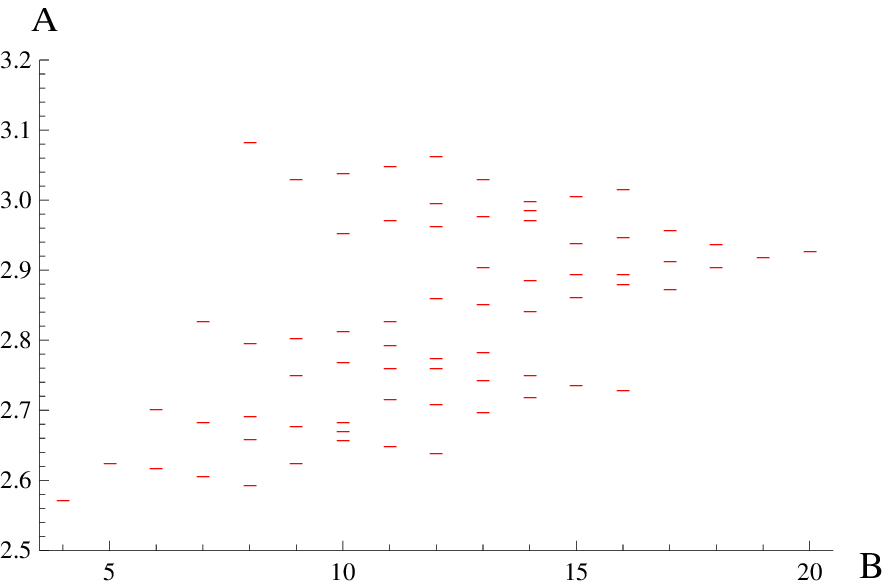}
   \end{minipage}
   \caption{Entanglement spectra of the $\nu=\half$ Laughlin (left) and
     HS (right) states for systems with 9 particles (or down-spins).
     The spectra consist {\it solely\/} of low level states; the
     entanglement gap is infinity.  The edge mode counting exactly
     matches the prediction from the CFT.  Both spectra possess exactly
     the same structure, while the HS spectrum - in absolute values of
     entanglement levels - is much flatter.}
 \label{hs}
 \vspace{-0pt}
 \end{figure}

\emph{(ii) Correspondence of Haldane-Shastry and Laughlin spectrum.}
As discussed in the manuscript, the Haldane-Shastry state and the
Laughlin state basically have the same entanglement spectrum, which
only differs in terms of the manifold where the states are defined.
Accordingly, the state counting exactly matches. However, the Laughlin
levels spread over a much bigger range of entanglement energy, while
the Haldane Shastry levels form an extremely flat band (Fig.~\ref{hs}).

\emph{(iii) Convergence of Lanczos algorithm.}
The biggest system size are computed by diagonalizing the spin
Hamiltonian in a hardcore boson model in momentum space. The
distribution of off-diagonal elements imposes a considerable challenge
for Lanczos optimization convergence to generate the eigenstate.
Exploiting all symmetries available in momentum space, the biggest
sizes considered ($N=22$) are matrices of dimension $3848879$ with
$\mathcal{O}(10^9)$ scattering elements. Convergence within a standard
error of $\mathcal{O}(10^{-8})$ for the Lanczos regeneration of the
eigenstate demanded up to $50000$ iterations.

$\phantom{ooo}$

\emph{(iv) Fitting procedures for the entanglement gap.}  There are
slightly different ways to define the entanglement gap, all of which
yield very similar results. The convention used in the paper is
\begin{equation}
\Delta := \text{Min}_{m \in [M^A]} \{\text{Min}_\xi \{ \xi^{\text{generic}}_{m}\} - \text{Max}_{\xi} \{ \xi^{\text{universal}}_{m}\}  \},
\end{equation}
i.e. the minimal gap between the highest low-energy universal level
and lowest high-energy generic level of all different momentum sectors
of $M^A$. Alternatively, one can define the entanglement gap globally
over all different momentum sectors $\Delta_{1} := \text{Min} \{
\xi^{\text{generic}}\} - \text{Max} \{ \xi^{\text{universal}}\}$. The
latter can be further generalized to a gap $\Delta_n$ where instead a
just taking the lowest generic level, we average over the lowest $n$
generic levels. If there were a spectral flow of only a single mode
closes the gap, the difference in $\Delta_{1}$ and $\Delta_{n}$ would
be able to detect it. For the cases considered in the manuscript,
however, we only observe a collective rise or lowering of levels sets
extending over all $M^A$ subsectors, which is why all these gap
definition nearly conincide in their result. 

\emph{(v) Spectral flow for the dimerization transition.}  A movie
(movie-j1j2.gif) of the spectral flow from the Heisenberg point to the
Majumdar Gosh point is added to the arxiv source files.

\end{document}